\documentclass[oldversion]{aa}  
\usepackage{graphicx}
\usepackage{epsfig}
\usepackage{natbib}
\usepackage{txfonts}
\begin{document}
   \title{
The metallicity of the most distant quasars
             \thanks{Based on data obtained at the VLT, through the ESO programs
			 074.B-0185 and 075.B-0487,
			 and at the Telescopio Nazionale Galileo.}
   }


   \author{
Y.~Juarez\inst{1}
          \and
		  R.~Maiolino\inst{2}
		  \and
          R.~Mujica\inst{1}
		  \and
		  M.~Pedani\inst{3}
		  \and
	      S.~Marinoni\inst{4}
		  \and
		  T.~Nagao\inst{5}
		  \and
		  A.~Marconi\inst{6}
		  \and
		  E.~Oliva\inst{7}
          }

   \institute{
             Instituto Nacional de Astrof\'{i}sica,
			 \'Optica y Electr\'onica,
             Puebla, Luis Enrique Erro 1, Tonantzintla, Puebla 72840, Mexico
		\and
   INAF-Osservatorio Astronomico di Roma, via di Frascati 33, 00040
              Monte Porzio Catone, Italy
         \and
	Large Binocular Telescope Observatory, University of Arizona,
	933 N. Cherry Ave., Tucson, AZ 85721-0065, USA
		\and
	INAF - Telescopio Nazionale Galileo, PO Box 565,
	38700 Santa Cruz de La Palma, Tenerife, Spain
		\and
	Research Center for Space and Cosmic Evolution, Ehime University,
	2-5 Bunkyo-cho, Matsuyama 790-8577, Japan
         \and
	Dipartimento di Astronomia, Universit\`a di Firenze,
	Largo E. Fermi 2, 50125 Firenze, Italy
         \and
	Osservatorio Astrofisico di Arcetri,
	Largo E. Fermi 5, 50125 Firenze, Italy
             }

   \date{Received ; accepted }

 
  \abstract
  {
We investigate the metallicity of the broad line region (BLR) of a sample of 
30 quasars in the redshift range 4$<$z$<$6.4, by using near-IR and
optical spectra. We focus on the ratio of the broad lines
(SiIV1397+OIV]1402)/CIV1549,
which is a good metallicity tracer of the BLR. We find that
the metallicity of the BLR is very high even in QSOs at z$\sim$6. The inferred metallicity
of the BLR gas is so high (several times solar) that metal ejection or mixing with lower metallicity
gas in the host galaxy is required to match the metallicities observed in local massive galaxies.
On average, the observed metallicity changes neither among quasars in the observed
redshift range 4$<$z$<$6.4, nor when compared with quasars at lower
redshifts. We show that the apparent lack of metallicity evolution
is a likely consequence of both the black hole-galaxy co-evolution and of selection effects.
The data also suggest a lack of evolution in the carbon abundance, even among z$>$6 quasars.
The latter result is puzzling,
since the minimum enrichment timescale of carbon is about 1~Gyr, i.e. longer than the age
of the universe at z$\sim$6.
}  

   \maketitle
%

\section{Introduction}
\label{sec_intro}

One of the important results from studies of high-redshift quasars
is the apparent lack of evolution in their emission properties. In particular,
the lack of evolution in the emission-line ratios is interpreted as a
lack of metallicity evolution among QSOs at different redshifts.
Previous tentative indications of metallicity evolution
with redshift were later ascribed to a metallicity dependence on the quasar
luminosity \citep{nagao06a}, which may mimic a dependence on redshift
in flux-limited samples. \cite{warner04} and \cite{shemmer04} suggest that the
metallicity-luminosity relation may actually result from a BH-metallicity dependence
or from a relation between accretion rate ($\rm L/L_{Edd}$) and metallicity.

At z$>$5 most of the emission lines used to constrain the BLR metallicity
are redshifted into the near-IR. As a consequence the investigation of the BLR metallicity
at z$>$5 has been limited to a small number of objects. Various authors have investigated
the iron abundance in z$\sim$6 QSOs by using near-IR spectra to measure the intensity
of the UV ``FeII-bump'' ($\lambda _{rest}\sim 2000-3050\AA$) relative to the MgII$\lambda$2789
line \citep{freudling03,maiolino03,iwamuro04,kurk07}, as a proxy of the Fe/$\alpha$-elements
abundance ratio.
No evolution was found in the ratio
FeII/MgII; however, since the FeII bump is a strong BLR coolant, its strength
depends on several other parameters and may
not depend primarily on the iron abundance.

\cite{pentericci02} investigated the metallicity in the BLR
of two QSOs at z$\sim$6 by measuring the ratio
of the CIV1549 and NV1240 emission lines, suggesting super-solar metallicities.
However, the calibration of the ratio NV/CIV as a tracer of gas metallicity
is mostly based on the assumption that nitrogen is a secondary element and that, therefore, its
abundance scales quadratically with metallicity, i.e.
$\rm N/H \propto (O/H)^2$. The latter assumption may, however, be an oversimplification and may
give unrealistically high metallicities, since the nitrogen abundance
actually depends on the detailed star formation history \citep{bresolin04,jiang08}.

\cite{nagao06a} show that the ratio
(SiIV$\lambda$1397+OIV]$\lambda$1402)/CIV$\lambda$1549 (hereafter (SiIV+OIV)/CIV)
is a tracer of the BLR metallicity,
the reason mainly being that the relative importance of CIV
as a coolant decreases with the BLR metallicity \cite{ferland96}.
This method has the advantage of not relying on the assumption
$\rm N/H \propto (O/H)^2$. Moreover, (SiIV+OIV) and CIV are well isolated features,
which do not require difficult deblending techniques and which can be measured even
in low-resolution spectra. \cite{jiang07} used near-IR spectra
to measure (SiIV+OIV)/CIV in five QSOs at z$\sim$6, and find no evidence of evolution
when compared with lower redshift QSOs.

In this paper we present measurements of the (SiIV+OIV)/CIV ratio in 30 QSOs at 4$<$z$<$6.4
obtained by means of near-IR spectra, with the goal of better investigating the metallicity
evolution of QSOs in the early Universe. In the following we adopt the following cosmological parameters:
   $H_{\rm 0}$=70 km s$^{-1}$ Mpc$^{-1}$, $\Omega_{\rm M}$=0.3, and $\Omega_{\rm \Lambda}$=0.7.


\section{Observations}

    We observed a sample of 30 high-redshift quasars ($4.0<z<6.4$) from the SDSS by means of
    near-IR and optical spectra covering at least the UV rest-frame emission lines 
    SiIV$\lambda$1397+OIV]$\lambda$1402 and CIV$\lambda$1549, but in most cases
	the spectra extend to $\rm \lambda _{rest}\sim 3000-4000\AA$. The original goal of most of
	the observations
	was to constrain the dust extinction in high-z QSOs. A more detailed description of the data and
	the results on the dust extinction will be given in Gallerani et al. (in prep.). Here we only
	focus on a byproduct, namely the evolution of the BLR metallicity based on the
	(SiIV+OIV)/CIV ratio.

    Observations were obtained both with the Italian Telescopio Nazionale Galileo (TNG) in Spain
    and with the Very Large Telescope (VLT)-ESO in Chile. Observations were performed in several
    observing runs from 2003 to 2005.
   The observations at the TNG were obtained with the Near Infrared Camera Spectrograph (NICS)
   mostly with the Amici prism
	to obtain spectra in the range 0.9-2.3 $\mu$m at R$\sim$75. This low-resolution mode is excellent for
	investigating the QSO continuum shape, but also for detecting broad emission lines.
	Some QSOs were observed again with the IJ grism to obtain
	0.9-1.45~$\mu$m spectra at R$\sim$500.
	Typical integration times ranged from $\sim$20 minutes to $\sim$3 hours.
	The observing strategy and data reduction are similar to those discussed in \cite{maiolino04}.

    The spectroscopic observations at ESO-VLT were done with the FORS2, along with the grism GRIS150I,
	to observe the range 6000-11000 \AA \ at R$\sim$300.
	These observations are mostly used to cover the
	short-wavelength part of some of the quasar spectra not properly sampled by the near-IR
	observations,
	but we also specifically observed a few quasars with no near-IR data with the specific aim of
	measuring the (SiIV+OIV)/CIV ratio.
    The total exposure times range from 30 to 60 minutes.
    For some of the $z < 5$ quasars observed with NICS, for which no FORS2 observations were
	available, we combine our near-IR spectra
	with optical data taken from \cite{anderson01}.


\begin{figure}[h]
\centering
\includegraphics[width=7.8cm]{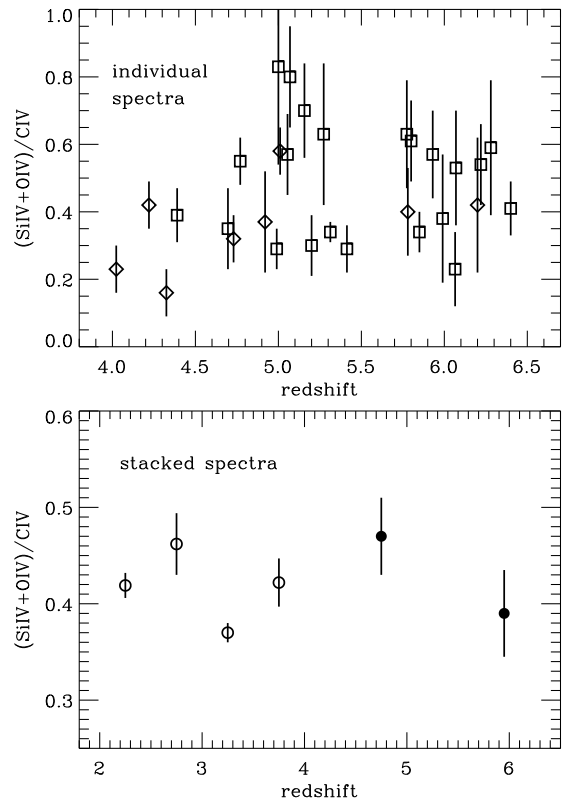}
\caption{
({\it Upper})  (SiIV+OIV)/CIV flux ratio as a function of redshift for quasars observed in this work. Squares
indicate non-BAL quasars, while diamonds indicate BAL quasars.
({\it Lower}) (SiIV+OIV)/CIV ratio inferred from quasar stacked spectra, both for the
high-redshift quasars presented here (filled symbols) and for the lower redshift quasars studied
in \cite{nagao06a} (empty symbols).}
\label{fig_z}
\end{figure}
\begin{figure}[h!]
\centering
\includegraphics[width=7.9cm]{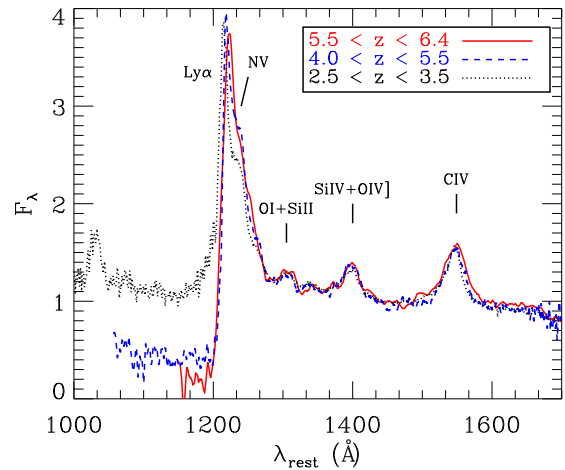}
\caption{Stacked spectra of quasars in different redshift bins (all normalized to $\rm \lambda = 1450 \AA$). The red solid
line and the blue dashed line are the stacked spectra from our sample in the redshift intervals
5.5$<$z$<$6.4 and 4.0$<$z$<$5.5,
respectively. The black dotted line is the SDSS quasars stacked spectrum in the redshift
interval 2.5$<$z$<$3.5.}
\label{fig_stack}
\end{figure}

\section{Results}


\begin{table}
\caption{(SiIV+OIV)/CIV measurements.}
\label{tab1}
\begin{tabular}{l l l l}
\hline
 Name             &  z    &  (SiIV+OIV)/CIV  & log$\lambda
L_{\lambda}^{\mathrm{a}}$      \\
\hline
SDSS J000239.4+255035 &  5.80   &  0.61$\pm$0.12                   & 46.93 \\
SDSS J000552.3-000656 &  5.85   &  0.34$\pm$0.06                   & 46.20 \\
SDSS J001714.66-100055.4$^{\mathrm{b}}$ &  5.01  &  0.58$\pm$0.07    & 46.56 \\
SDSS J012004.82+141108.2$^{\mathrm{b}}$ &  4.73  &  0.32$\pm$0.03    & 46.09 \\
SDSS J015642.11+141944.3$^{\mathrm{b}}$ &  4.32  &  0.16$\pm$0.03    & 46.62 \\
SDSS J023137.6-072855 &  5.41  &  0.29$\pm$0.07                   & 46.56 \\
SDSS J023923.47-081005.1$^{\mathrm{b}}$ &  4.02   &  0.23$\pm$0.04    & 46.58 \\
SDSS J033829.3+002156 &  5.00   &  0.97$\pm$0.29                   & 46.46 \\
SDSS J075618.1+410408 &  5.07   &  0.80$\pm$0.15                   & 46.53 \\
SDSS J083643.8+005453 &  5.80   &  0.63$\pm$0.16                   & 46.97 \\
SDSS J085210.89+535948.9$^{\mathrm{b}}$ &  4.22   &  0.42$\pm$0.04  & 46.53 \\
SDSS J095707.67+061059.5 &  5.16  &  0.70$\pm$0.14          & 46.65 \\
SDSSp J102119.16-030937.2 &  4.70  &  0.35$\pm$0.12         & 46.58 \\
SDSS J103027.1+052455 &  6.28   &  0.59$\pm$0.20                   & 46.68 \\
SDSS J104433.04-012502.2$^{\mathrm{b}}$ & 5.78 &    0.40$\pm$0.13			& 46.88 \\
SDSS J104845.05+463718.3$^{\mathrm{b}}$ & 6.20 &    0.42$\pm$0.20			& 46.81 \\
SDSS J114816.6+525150 &  6.40   &  0.41$\pm$0.08                   & 46.95 \\
SDSS J120441.7-002150 &  5.05  &  0.57$\pm$0.12                   & 46.63 \\
SDSSp J120823.8+001028 &  5.27  &  0.63$\pm$0.21                  & 46.07 \\
SDSS J130608.2+035626 &  5.99   &  0.38$\pm$0.19                   & 47.32 \\
SDSS J141111.3+121737 &  5.93   &  0.57$\pm$0.13                   & 46.58 \\
SDSS J160254.2+422823 &  6.07   &  0.53$\pm$0.17                   & 46.90 \\
SDSS J160320.89+072104.5 &  4.39  &  0.39$\pm$0.08           & 46.85 \\
SDSS J160501.21-011220.6$^{\mathrm{b}}$ & 4.92   & 0.37$\pm$0.15 				& 46.46  \\
SDSS J161425.13+464028.9 &  5.31  &  0.34$\pm$0.03          & 46.62 \\
SDSS J162331.8+311201 &  6.22   &  0.54$\pm$0.12                   & 46.54 \\
SDSS J162626.50+275132.4 &  5.20   &  0.30$\pm$0.09          & 46.94 \\
SDSS J163033.9+401210 &  6.06  &  0.23$\pm$0.11                   & 46.38 \\
SDSS J220008.7+001744 &  4.77   &  0.55$\pm$0.07                   & 46.58 \\
SDSS J221644.0+001348 & 4.99    &  0.29$\pm$0.06                   & 46.18 \\
\hline
\end{tabular}
$^{\mathrm{a}}$ Rest-frame continuum luminosity at 1450\AA, in units of erg~s$^{-1}$.\\
$^{\mathrm{b}}$ BAL quasars.
\end{table}

\begin{figure}
\centering
\includegraphics[width=8cm]{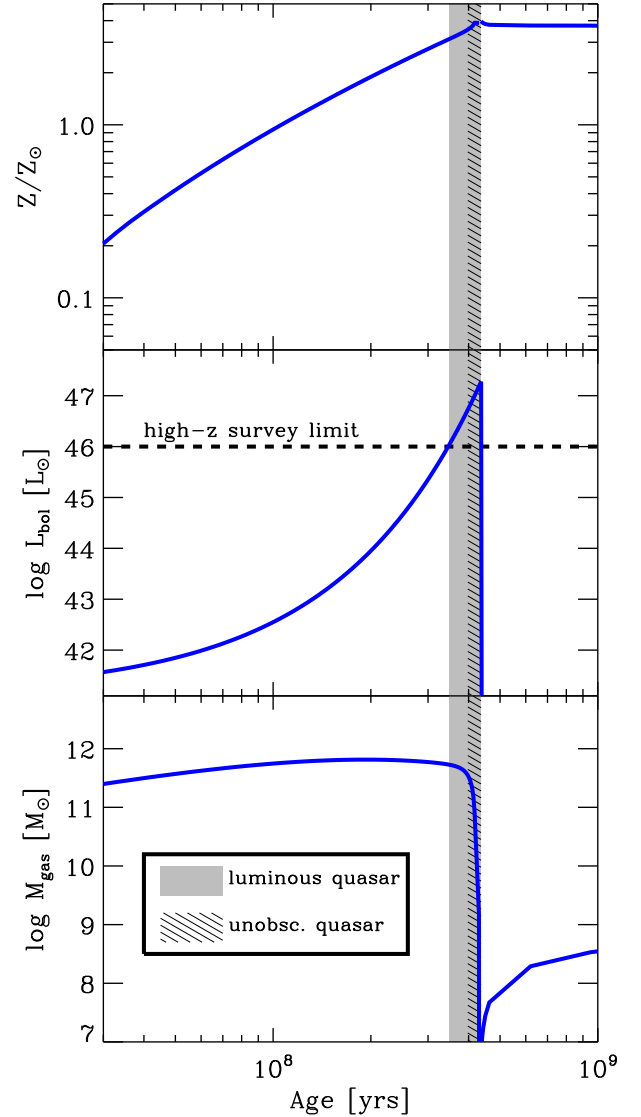}
\caption{Evolution of the gas metallicity (top), bolometric quasar luminosity (middle) and total gas mass
(bottom) in the model by \cite{granato04}.
The horizontal dashed line shows the approximate minimum bolometric luminosity detectable at high-z in the SDSS quasar survey.
The shaded area shows the epoch when a high-z quasar is luminous enough to be detected in the SDSS survey (if unobscured).
The hatched area is the epoch when the quasar has ejected more than half of the gas mass, and it is likely unobscured and
therefore detectable in optical surveys. By the time the quasar has reached the luminous, ``unobscured'' epoch the
gas is already highly enriched.
}
\label{fig_model}
\end{figure}


While individual spectra will be shown in a forthcoming paper (Gallerani et al., in prep.), in this
letter we report the measurements of the (SiIV+OIV)/CIV ratio.
To measure emission line fluxes, these low resolution quasar spectra were fitted with a model
consisting of a power-law continuum and Gaussian profiles to
model the emission lines. The power-law continuum was fitted using small spectral regions free of
emission features (1345--1355\AA,1445--1455\AA,1695-1704\AA,1973--1983\AA).
The SiIV+OIV] blend is unresolved and is fitted with
a single Gaussian.
Eight of our sources are broad absorption line (BAL) quasars. In these cases the measurement of
the emission line flux is more complex and uncertain because of the absorption troughs, which
are avoided when fitting both the continuum and the emission lines. 
Table~\ref{tab1} lists the (SiIV+OIV)/CIV ratio for all quasars in our sample, along
with their redshift and continuum luminosity ($\lambda L_{\lambda}$ at 1450\AA).
Figure \ref{fig_z} (top panel) shows the measured (SiIV+OIV)/CIV ratios
as a function of redshift. According to the calibration given in \cite{nagao06a}, the observed ratios
correspond to metallicities that are several times solar.
No correlation is observed between the emission line ratio (hence metallicity) and redshift.
Figure~\ref{fig_stack} shows the average
spectra in the redshift intervals 4.0$<$z$<$5.5 and 5.5$<$z$<$6.4 (obtained by first normalizing
individual spectra to the continuum at $\rm \lambda 1450\AA$, and avoiding BAL quasars),
which are remarkably similar\footnote{Small differences in Ly$\alpha$ and NV are due to different absorption
of the IGM and to different spectral resolution of the spectra.}, again highlighting any lack of evolution.

Our results can be compared with the line ratios and metallicities inferred at even lower redshifts, as obtained
by \cite{nagao06a}. It is important to use quasars with similar luminosities, since
quasar metallicities show a strong correlation with luminosity. Given that
flux limited surveys probe different luminosities at different redshifts, this may introduce
an apparent redshift evolution. Therefore, from the 2$<$z$<$4 templates in \cite{nagao06a}, we select those in
the highest luminosity bin ($\rm -28.5 > M_B > -29.5$), which match the quasar luminosities in our sample
(which are also restricted to a relatively narrow range).
In Fig.~\ref{fig_stack} the average spectra of z$>$4 quasars are compared
with the average spectrum of luminous QSOs at 2.5$<$z$<$3.5, which are all very similar.
The lack of any evolution with redshift can be also appreciated in the bottom
panel of Fig.\ref{fig_z}, which shows
the ratio (SiIV+OIV)/CIV measured in average quasar spectra at different redshifts (matched in
luminosity).

	\section{Discussion}

\subsection{The quasar metallicity ``evolution''}

The apparent lack of evolution observed in Figs.~\ref{fig_z}--\ref{fig_stack} should not be interpreted as a lack of evolution
of the BLR metallicity in individual AGNs. Indeed, Fig.~\ref{fig_z} shows the
average metallicity of the BLR in quasars that
are accreting at the given redshift, but does {\it not} trace the evolutionary path of individual quasars. The apparent lack of
evolution in the BLR metallicity observed in Fig.~\ref{fig_z} likely results from a combination of the BH-galaxy
co-evolution and selection effects. Indeed, to cross the detection threshold of the SDSS magnitude-limited survey,
high-redshift quasars must have high luminosities, hence (even if accreting at the Eddington limit) high black hole masses.
Most models predict that high black hole masses must have been accompanied
by the formation of a massive host galaxy
\citep[e.g.][]{granato04,dimatteo05,hopkins08,li07}, which would result into
the local $\rm M_{BH}-M_{spheroid}$ relationship.
Therefore, by the time a quasar at any redshift is detectable in a magnitude-limited
survey, its host galaxy must have evolved significantly and enriched its ISM significantly.
The quasar feedback is another evolutionary effect that may yield to observational biases resulting in an apparent
lack of metallicity evolution. Indeed, according to many models, during the early phases, when the host galaxy is still
metal poor, the accreting black hole is embedded within the dusty ISM, and therefore difficult to detect in optical surveys.
Only during the late evolutionary phases, when the galaxy is already metal rich, the quasar develops winds powerful enough
to expel large quantities of gas and dust, so that the quasar becomes visible to optical observations.

To show these combined effects in a more quantitative way, we exploited the quasar-galaxy evolutionary models
reported in \cite{granato04}.
In Fig.~\ref{fig_model} we show the evolution of the gas metallicity
in a massive galaxy forming in a dark halo of $\rm M_{halo}=10^{13.2}~M_{\odot}$,
along with the evolution of the quasar bolometric luminosity and the total gas mass.
While the host galaxy forms stars, the gas metallicity increases. At the same time the black hole accretion increases, yielding
an increasingly high quasar luminosity. At z$>$5 the SDSS magnitude limit translates into a minimum quasar bolometric luminosity
of about $\rm 10^{46}~erg~s^{-1}$ for detection. Fig.~\ref{fig_model} shows that by the time the quasar reaches such high
luminosity (shaded region),
the metallicity in the host galaxy is already higher than $\rm \sim 3~Z_{\odot}$. At this time the model expects the
quasar to be probably still embedded in gas and dust, although the quasar has already developed a wind expelling gas.
It is not easy to identify the stage when the quasar becomes unobscured and detectable in optical surveys,
since it depends on the detailed distribution of dust.
We assume that the quasar becomes optically visible when more than half of the gas mass has been expelled. This is a somewhat
arbitrary assumption, but we note that it is roughly consistent with what obtained by \cite{lapi06}, who identify the epoch
of unobscured accretion as starting a few $\rm 10^7~yrs$ before the epoch of maximum luminosity.
In Fig.~\ref{fig_model} we note that by the time the quasar is unobscured (hatched region),
the gas in the host galaxy has already reached a metallicity of about $\rm \sim 4~Z_{\odot}$.
Summarizing, the co-evolution of black holes and galaxies, combined with
observational selection effects (mostly in optical surveys), naturally explains the finding that unobscured
quasars of a given luminosity appear to have on average the same metallicity at any redshift.

\subsection{The extreme metallicities in the BLR}

According to \cite{nagao06a}, the ratio (SiIV+OIV)/CIV$\sim$0.4 observed in the stacked
spectrum of the most distant quasars corresponds to a gas metallicity of $\rm \sim 7~Z_{\odot}$. Such huge metallicities
were also inferred by \cite{nagao06a} based on a much wider set of lines of lower redshift quasars.
These high gas metallicities are not unrealistic. Indeed,
the BLR is a small nuclear region (less than a few pc in the most luminous QSOs) with masses of a few times 
$\rm \sim 10^4~M_{\odot}$, which can be enriched {\it in situ}
to super-solar metallicities within less than $\rm 10^8 yrs$ by having
a supernova explosion less often than every $\rm 10^4 yrs$.
More troublesome is that such high metallicities
are not found in the stellar population of local massive galaxies, not even in their central regions.
Therefore, the high-metallicity gas observed in the BLR of high-z quasars must either be expelled (as in
quasar feedback models) or diluted
by lower metallicity gas in the host galaxy before forming stars.

A related issue is how much the BLR metallicity is representative of the gas metallicity in the host galaxy.
Investigating this issue would require accessing metallicity diagnostics that directly trace the quasar
host galaxies. Host galaxies have been probed either through the narrow emission lines \citep[only observable in obscured
QSOs, e.g.][]{nagao06b} or
through the associated narrow absorption lines \citep[e.g.][]{dodorico04}, but only at intermediate redshifts (z$<$3).
It was found that the metallicity of the host galaxy is typically lower than observed in the BLR, but still super-solar,
and again without any evidence of evolution with redshift. No direct measurements are currently available
for the metallicity in the host galaxy of z$\sim$6 quasars.
However, the detection of huge quantities of dust \citep{beelen06}
suggests that even the host of these quasars are heavily enriched.

\subsection{The carbon problem}

Relative elemental abundances are much more difficult to infer. However, the observational result that
the ratio (SiIV+OIV)/CIV does not evolve suggests that the abundance of carbon relative
to silicon and oxygen also does not evolve significantly. Large carbon abundances in the hosts
of the most distant quasars is also suggested by the detection of strong [CII]158$\mu$m and CO lines
in the most distant quasars \citep{maiolino05,bertoldi03}. Carbon is mostly produced by AGB stars and
planetary nebulae,
most of which evolve on long timescales. Although the first AGB stars appear as soon as $\sim$50~Myr
after the onset of star formation, the bulk of carbon production occurs after about 1~Gyr, yielding a delayed enrichment
with respect to $\alpha$-elements (e.g. oxygen) that are promptly produced by
SNII\footnote{Silicon is also produced by
SNIa and therefore is also probably enriched with a delay. However, \cite{nagao06a} find in their higher resolution spectra
that probably most of the emission in the SiIV+OIV blend is mostly due to OIV (see Tables 3--7).}. At z$>$6 the age of
the universe is less than 1~Gyr, hence stellar evolution may fall short of the time needed
to produce the observed large carbon abundance.
We note that this issue is independent of the size and mass of the BLR, making it just a pure timescale problem.
Tackling this issue requires a more accurate determination of the carbon abundance, which may come from future high
spectral resolution optical/near-IR observations or from future
submm observations of far-IR fine structure lines \citep{maiolino08}.

\begin{acknowledgements}
We are grateful to L.~Silva and G.L.~Granato for providing the electronic form of their models and for useful comments.
      This work was partially supported by the CONACyT project \#45948, by CONACyT grant for Ph.D studies \#157846, by
	  INAF and by ASI 
through contract ASI-INAF I/016/07/0. 
\end{acknowledgements}

\end{document}